\documentclass[aps,prb,preprint,superscriptaddress,showpacs]{revtex4-1}
\usepackage{graphicx}
\usepackage{xcolor}
\usepackage{rotating}
\usepackage{hyperref}
\usepackage{array}
\newcolumntype{C}[1]{>{\centering\let\newline\\\arraybackslash\hspace{0pt}}m{#1}}
\renewcommand{\figurename}{Fig.}
\begin{document}

%\preprint{}

\title{Symmetry protected 1D chains in mixed-valence iron oxides}

\author{Denis~M.~Vasiukov}
\email[]{vasiukov.dv@gmail.com}

\affiliation{Division of Synchrotron Radiation Research, Department of Physics, Lund University, Box 118, Lund 221 00, Sweden}
\author{Ghanashyam~Khanal}
\affiliation{Department of Physics and Astronomy, Rutgers University, Piscataway, New Jersey 08854, USA}
\author{Ilya~Kupenko}
\affiliation{Institut f\"{u}r Mineralogie, University of M\"{u}nster, Corrensstr. 24, 48149 M\"{u}nster, Germany}
\author{Georgios~Aprilis}
\affiliation{ESRF, The European Synchrotron, 71 Avenue des Martyrs, CS40220, 38043 Grenoble Cedex 9, France}
\author{Sergey~V.~Ovsyannikov}
\affiliation{Bayerisches Geoinstitut, Universit\"{a}t Bayreuth, Universit\"{a}tsstr. 30, D-95447 Bayreuth, Germany}
\author{Stella~Chariton}
\affiliation{Bayerisches Geoinstitut, Universit\"{a}t Bayreuth, Universit\"{a}tsstr. 30, D-95447 Bayreuth, Germany}
\author{Valerio~Cerantola}
\affiliation{ESRF, The European Synchrotron, 71 Avenue des Martyrs, CS40220, 38043 Grenoble Cedex 9, France}
\author{Vasily~Potapkin}
\affiliation{Bayerisches Geoinstitut, Universit\"{a}t Bayreuth, Universit\"{a}tsstr. 30, D-95447 Bayreuth, Germany}
\author{Aleksandr~I.~Chumakov}
\affiliation{ESRF, The European Synchrotron, 71 Avenue des Martyrs, CS40220, 38043 Grenoble Cedex 9, France}
\author{Leonid~Dubrovinsky}
\affiliation{Bayerisches Geoinstitut, Universit\"{a}t Bayreuth, Universit\"{a}tsstr. 30, D-95447 Bayreuth, Germany}
\author{Kristjan~Haule}
\affiliation{Department of Physics and Astronomy, Rutgers University, Piscataway, New Jersey 08854, USA}
\author{Elizabeth~Blackburn}
\affiliation{Division of Synchrotron Radiation Research, Department of Physics, Lund University, Box 118, Lund 221 00, Sweden}

\date{\today}

\begin{abstract}
During the last decade of high-pressure research a whole new series of iron oxides was discovered, like~Fe$_4$O$_5$, Fe$_5$O$_6$, Fe$_7$O$_9$~etc.~[\onlinecite{lavina2011discovery,lavina2015unraveling,sinmyo2016discovery,bykova2016structural,ishii2018synthesis,koemets2021chemical}], featuring closely related structures with arrays of one-dimensional (1D) chains of trigonal prisms embedded between slabs of octahedra. Here, we develop a unified approach to the series based on a specific crystallographic generation mechanism which predicts the structures of these oxides and naturally classifies them in terms of the slab cycle. When including magnetic interactions, we show that the 1D chains have a symmetry protection against magnetic perturbations from the iron ions in the slabs, and that the slab size determines the type of magnetic order, which is either ferromagnetic or antiferromagnetic. Dynamical mean-field theory calculations reveal the orbitally selective Mott state of the Fe ions and tendency of conductivity to low-dimensional behavior with particular enhancement along the 1D chains. Across the series, the decoupling of the chains increases, and so with the inherent charge ordering of the slabs, these structures have the potential to allow experimental realization of the model system of coupled 1D wires~[\onlinecite{vishwanath2001two,kane2002fractional,teo2014luttinger,iadecola2016wire,meng2015coupled}]. We point out the possibility to stabilize these compounds in the thin-film form that, together with a wide range of possible ionic substitutions and fact that these compounds are recoverable at ambient pressure, makes them a very promising platform to engineer physical systems with interesting magnetotransport phenomena, as corroborated by the recent discovery of quantum Hall effect in ZrTe$_5$~[\onlinecite{tang2019three}].

%Iron oxides are normally Mott insulators with the only notable exception of magnetite (Fe$_3$O$_4$), which has good electrical conductivity due to the itinerant mixed-valence state of iron.
\end{abstract}

%\pacs{75.30.Wx, 76.80.+y, 61.50.Ks}
%\keywords{}

\maketitle


%\paragraph*{\textbf{Introduction}}
%Iron oxides have accompanied mankind's development throughout the ages, from the use of ochre in the late Stone Age in the celebrated cave paintings found around the world, to the extraction of iron for tools in the Iron Age. Their magnetic effects led not only to applications which revolutionized our daily life, like compasses or magnetic recording, but also played a key role in gaining understanding of magnetic phenomena on the very fundamental level that culminated in the pioneering neutron diffraction studies of Shull {\it et al.} [\onlinecite{shull1951antiferro,shull1951feri}] confirming N\'eel's concept of antiferromagnetism and ferrimagnetism \cite{neel1932influence,neel1936proprietes,neel1948proprietes}.

%Due to their high abundance and high critical temperatures, iron oxides were proposed to be responsible for deep mantle magnetic anomalies, challenging the concept of the non-magnetic mantle of our planet\cite{kupenko2019magnetism}. The presence of these phases in the Earth's mantle and their potential influence on the mantle properties are now the focus of numerous studies in geoscience\cite{myhill2016p,sinmyo2019effect,kupenko2019magnetism,Anzolini2020Evidence}.

Of the many new high-pressure iron oxides discovered during the last decade, the set consisting of Fe$_4$O$_5$, Fe$_5$O$_6$, HP-Fe$_3$O$_4$ etc.~[\onlinecite{lavina2011discovery,lavina2015unraveling,sinmyo2016discovery,bykova2016structural,ishii2018synthesis,koemets2021chemical}] is particularly special. The oxides in this set share a common packing motif and can be united into a $n$FeO$\cdot m$Fe$_2$O$_3$ homologous series\cite{bykova2016structural}. %\textcolor{brown}{[These compounds are of interest not only for geoscience[refs], but also for solid state physics and material science since]}
Many members of this series are recoverable at ambient conditions after high-pressure synthesis and feature iron ions in mixed-valent states\cite{ovsyannikov2018pressure,sinmyo2016discovery,maitani2022electrical}, resembling magnetite (Fe$_3$O$_4$), where this electronic state results in the famous metal-insulator Verwey transition at low temperatures\cite{verwey1939electronic}, with formation of a peculiar charge-ordered ground state with trimer units of iron ions\cite{senn2012charge,baldini2020discovery}. Indeed, similar Verwey-type transitions were found in Fe$_5$O$_6$~[\onlinecite{ovsyannikov2020room}] and Fe$_4$O$_5$~[\onlinecite{ovsyannikov2016charge,ovsyannikov2018pressure}].

In this paper we focus on an overlooked feature of these compounds, namely that part of the iron ions is arranged in arrays of symmetry protected 1D chains. We use both experimental and computational methods to explore these arrays and argue that, under certain plausible conditions, they could host coupled one-dimensional (1D) wires. Such systems have been extensively studied in theory, but have proven difficult to realize experimentally\cite{vishwanath2001two,kane2002fractional,teo2014luttinger,iadecola2016wire,meng2015coupled}. These models yield a broad spectrum of fascinating states, from non-Fermi liquid\cite{vishwanath2001two} to a zoo of topological phases\cite{kane2002fractional,teo2014luttinger,iadecola2016wire}, including fractional quantum Hall states, in the case of conductive wires (\emph{viz}~Luttinger liquids), and spin liquids in the case of insulating wires\cite{meng2015coupled}.

\begin{figure*}
\includegraphics[width=\linewidth, keepaspectratio=true]{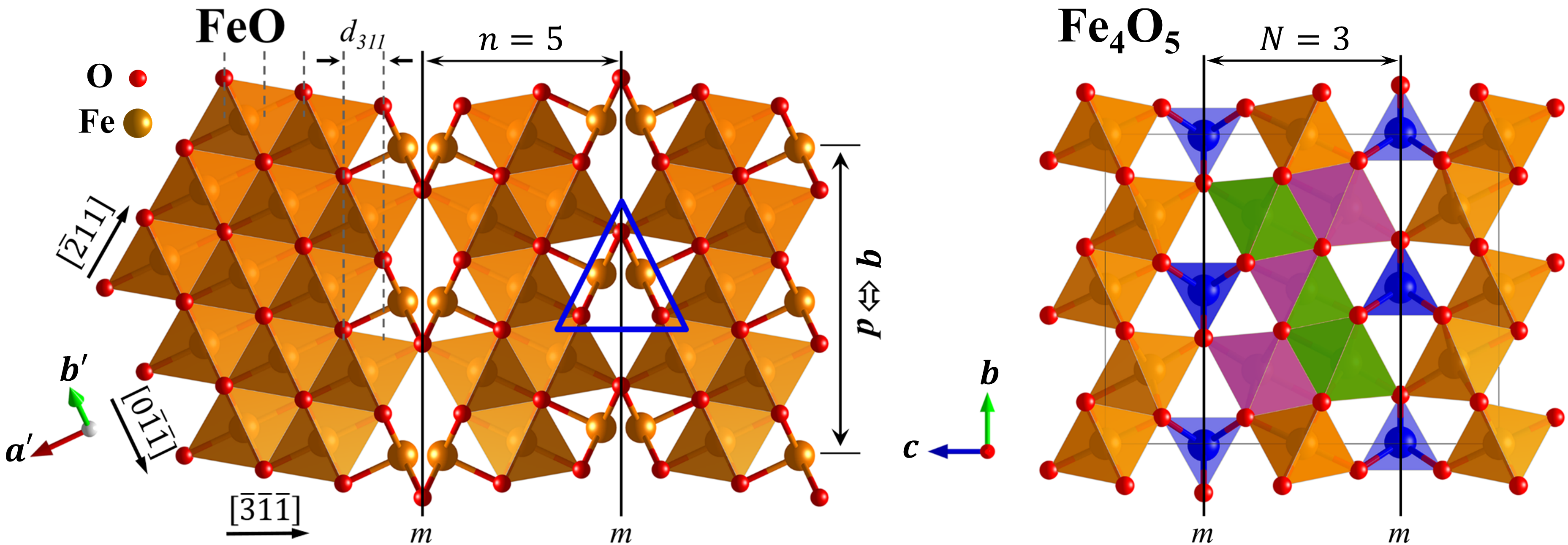}
\caption{\label{Figure_1}The $\{311\}$ tropochemical cell-twinning in the homologous series of iron oxides as exemplified by the Fe$_4$O$_5$ structure. On the left, the FeO structure is projected along~$[0\bar{1}1]$. Regularly applying a mirror plane parallel to~(311) at the oxygen sheet at $5d_{311}^{\text{FeO}}$~intervals (slab cycle 3), we derive a structure with a new unit cell. Upon this operation, pairs of iron atoms, as outlined by the blue triangle, are brought into proximity across twinning planes. The last step is the coalescence of these pairs into a single iron atom that leads to a new crystallographic site at the twinning plane with a trigonal prismatic oxygen environment, as marked by the blue colour in the Fe$_4$O$_5$ structure on the right. These trigonal prisms form infinite 1D chains along the $a$ axis in the resultant structure. The slab width $N$ can be readily determined using either of two equivalent octahedral chains highlighted by the violet and green cages in the Fe$_4$O$_5$ structure. For this particular slab cycle the twinning plane is preserved as a symmetry element in the generated structure. The periodicity $p$ in the plane of projection of FeO is equivalent to the $b$ lattice constant of a new unit cell. Hereafter we use orange and blue colors to designate the octahedral and trigonal prismatic sites respectively.}
\end{figure*}

%Our current understanding of physical properties of these novel oxides is very limited.  Experimentally, only a very small amount of these materials or tiny single crystals can be produced by high-pressure synthesis techniques and some of them are not recoverable at ambient conditions. Altogether this makes their comprehensive investigation challenging and seriously limits  available experimental techniques.

%\textcolor{brown}{[Here we develop a unified crystallographic approach to these compounds. Based on it and M\"{o}ssbauer spectroscopy data, we determine a generic magnetic structure for the charge-averaged state in these oxides which is used for the state-of-the-art dynamical mean-field theory (DMFT) calculations of electronic band structures and analysis of transport properties.]}

%\paragraph*{\textbf{Crystallography of the $n$FeO$\cdot m$Fe$_2$O$_3$ series}}
We begin with the new insight that stoichiometric variation in the $n$FeO$\cdot m$Fe$_2$O$_3$ series resembles the well-known Magn\'{e}li phases\cite{schwingenschlogl2004vanadium}. This motivated us to investigate the underlying crystallographic mechanism responsible for generating the variety of structures seen in this homologous series. There is a close link between the $n$FeO$\cdot m$Fe$_2$O$_3$ series and the ambient pressure $n$PbS$\cdot m$Bi$_2$S$_3$ homologous series, with isostructural correspondence between Fe$_5$O$_6$ and the mineral lillianite Pb$_3$Bi$_2$S$_6$, and between Fe$_5$O$_7$ and PbBi$_4$S$_7$, the so-called V1 phase~\cite{takeuchi1997tropochemical}. This resemblance reveals the common crystallographic generative mechanism behind both series. In the case of the $n$PbS$\cdot m$Bi$_2$S$_3$ series, it was studied in detail and named tropochemical cell-twinning\cite{takeuchi1997tropochemical}. We illustrate this mechanism in Figure~\ref{Figure_1} by way of deriving the Fe$_4$O$_5$ structure from w\"{u}stite (FeO).

The starting point is the parent rocksalt structure of w\"{u}stite. If a mirror plane is applied parallel to the \{311\}~plane at a sheet of oxygen atoms, we generate a twin. Consecutively applying this operation with some periodic spacing (multiple of~$d_{311}^{\text{FeO}}$), we get a new unit cell with identical slabs of the FeO$_6$ octahedra between twinning planes. As seen from Figure~\ref{Figure_1}, some iron atoms are brought together at the twinning plane upon this procedure. Merging these atoms gives a new iron site with a trigonal prismatic oxygen environment. The trigonal prisms share common triangular facets and form an array of 1D chains along the~$a$~axis separating the octahedral slabs (Fig.~\ref{Figure_1}). This also eliminates iron atoms from the structure, modifying the stoichiometry.

Depending on the spacing between twinning planes, one obtains structures with different slab widths. There are two equivalent ways to define this slab size: (\emph{i}) the number, $n$, of $d_{311}^{\text{FeO}}$ spacings~--~in other words the number of the oxygen sheets per slab~--~and (\emph{ii}) the number, $N$, of octahedra in either of the two octahedral chains along $\langle\bar{2}11\rangle$ and $\langle0\bar{1}\bar{1}\rangle$ directions in the initial w\"{u}stite structure. Hereafter we will use the latter notation, but there is a simple conversion between them, namely $n=N+2$.

The $\{311\}$ cell-twinning mechanism considered here can eliminate up to one third of the cations from the initial rocksalt structure given the smallest slab width of one octahedron, which corresponds to the $\eta$-Fe$_2$O$_3$ phase~\cite{bykova2016structural}. Variation in stoichiometry leads to changes in the average cation valence state, starting from pure~+2 in FeO to pure~+3 in $\eta$-Fe$_2$O$_3$, with the formation of numerous mixed-valence compounds at intermediate stoichiometries.

In Extended Data Table~I, the known members of the $n$FeO$\cdot m$Fe$_2$O$_3$ series are collected. It is natural to classify these oxides according to the slab cycle, or sequence of slabs, which entirely determines the symmetry and crystallographic positions in the generated structures. Most of the compounds in Extended Data Table~I can be assigned to two groups, the~$N$-family and $N$,$N$+1-family. The $N$-family has the simplest cycle, with only one type of slab, and crystallizes in an orthorhombic space group~$Cmcm$~(\#63). The mirror plane perpendicular to the~$c$~axis in this family is the initial twinning plane as illustrated in Figure~\ref{Figure_1}. There is a subtle crystallographic difference between compounds with even and odd~$N$ that has important consequences for the magnetic structure (see Supplementary Materials for details).

The $N$,$N$+1-family contains slabs of two different widths differing by one octahedron. This slab cycle does not allow preservation of the twinning plane as a symmetry element in the resulting structures. Instead, only symmetry elements relevant to the~$a$~axis of $Cmcm$~space group are preserved, and the unit cell becomes monoclinic with space group~$C2/m$~(\#12).

Generally speaking the slab cycle can be arbitrarily complex, as seen in the phase-V~subseries~of the lillianite structure\cite{takeuchi1997tropochemical}. The first sign of this complexity in the iron oxides is the recently discovered Fe$_7$O$_{10}$~[\onlinecite{koemets2021chemical}], with a trinomial slab cycle 1,1,2 which contains two different twinning planes between the 1,1 and 1,2 slabs, where only the first one is preserved as a mirror plane in the resulting structure. New iron oxides with more complex slab cycle are likely to be discovered in the future.

\begin{turnpage}
\begin{figure*}
\includegraphics[width=\linewidth, keepaspectratio=true]{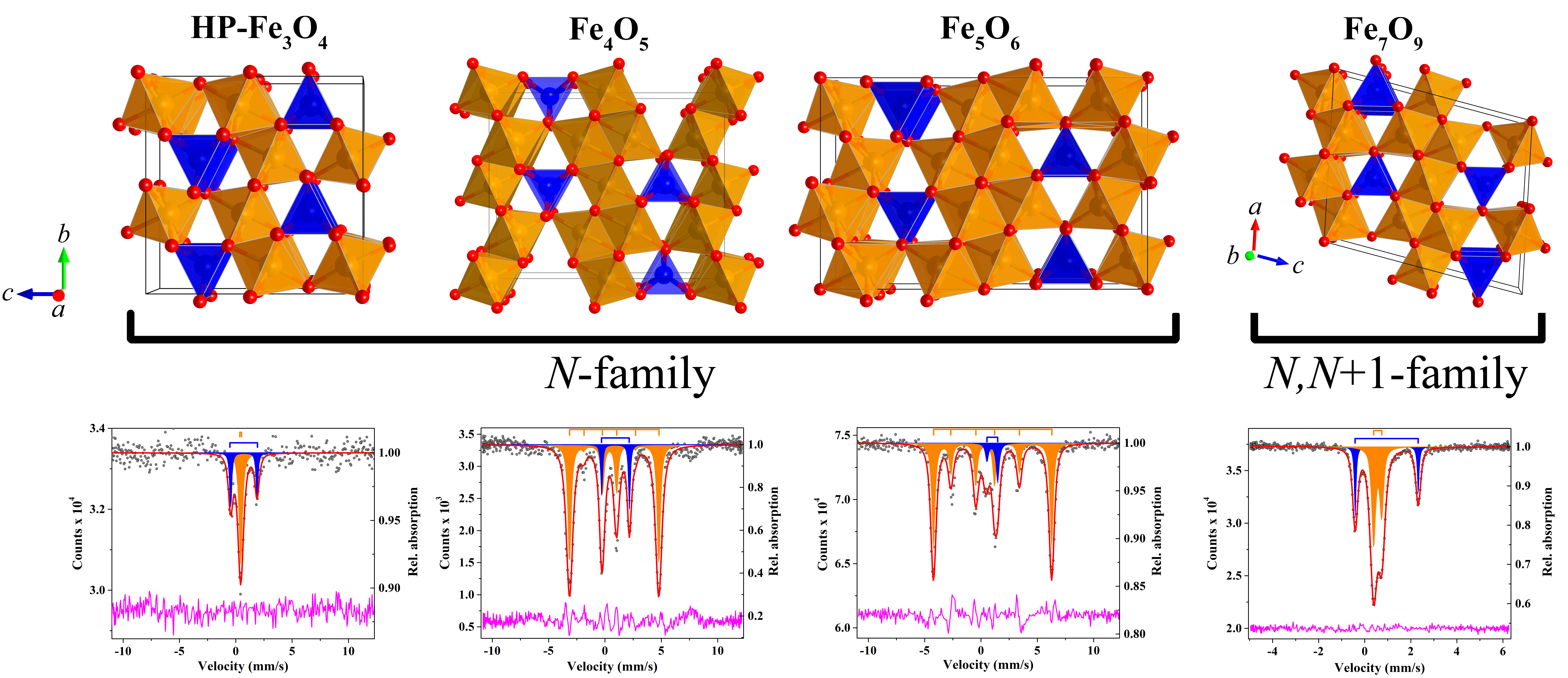}
\caption{\label{Figure_2} M\"{o}ssbauer spectra of several members of the $n$FeO$\cdot m$Fe$_2$O$_3$ homologous series at room temperature. The blue and orange components correspond to iron ions in the trigonal prismatic and octahedral environment, respectively. The pink line is the fit residual. In the paramagnetic region all of these oxides show similar spectra consisting of two distinct doublets corresponding to the iron ions in different oxygen environment, here exemplified by HP-Fe$_3$O$_4$ and Fe$_7$O$_9$. The spectra of the Fe$_4$O$_5$ and Fe$_4$O$_5$ oxides show an unexpected coexistence of the magnetic sextet from the octahedral slab with the paramagnetic doublet of the trigonal prismatic site. This means that magnetic ordering in the octahedral slab does not influence iron ions in the chains of trigonal prisms. The spectrum of HP-Fe$_3$O$_4$ was collected at 20~GPa whereas other spectra were acquired at ambient pressure; the spectrum of Fe$_7$O$_9$ was already published in~[\onlinecite{sinmyo2016discovery}].}
\end{figure*}
\end{turnpage}

%\paragraph*{\textbf{Results of M\"{o}ssbauer spectroscopy}}
%\textcolor{brown}{For such iron-containing compounds the use of M\"{o}ssbauer spectroscopy is particularly beneficial as it can provide unique information about behavior of each independent crystallographic iron site. We carried out M\"{o}ssbauer experiments on Fe$_3$O$_4$, Fe$_4$O$_5$, Fe$_5$O$_6$ and Fe$_7$O$_9$ using Synchrotron M\"{o}ssbauer Source (SMS) at the Nuclear Resonance beamline (ID18) at ESRF.}

The representative M\"{o}ssbauer spectra of the studied oxides are shown in Fig.~\ref{Figure_2}. Their basic feature is the presence of two distinct components. One belongs to the octahedral slab while the other corresponds to the chains of trigonal prisms. The data show that in all recoverable oxides, namely Fe$_4$O$_5$, Fe$_5$O$_6$ and Fe$_7$O$_9$, the chains are exclusively populated by Fe$^{2+}$ ions at ambient conditions. This observation allows us to estimate the average oxidation state of the Fe ions in the slabs and the isomer shifts of the slab components are indeed in a good agreement with the predicted values (Extended Data Table~II).
%This corollary, together with the fact that octahedra in the slabs are interconnected via common edges that favour direct $d$--$d$ orbital interactions between adjacent iron ions, leads to the formation of the mixed-valent state that is responsible for the good electrical conductivity observed in experiments\cite{ovsyannikov2016charge,ovsyannikov2020room}.

An intriguing feature of the spectra in~Fig.~\ref{Figure_2} is the apparent coexistence of the magnetically ordered iron ions in the slabs with the disordered iron ions in the chains of Fe$_4$O$_5$ and Fe$_5$O$_6$. Accordingly, we deduce that in both compounds slabs and chains have two different ordering temperatures and that the magnetic ordering of the 1D chains is independent from that of the slabs. This peculiar magnetic behavior was also observed in the isostructural compounds MnFe$_3$O$_5$ and CoFe$_3$O$_5$~[\onlinecite{hong2018MnFe3O5,hong2018CoFe3O5}] and it is confirmed to be stable by our theoretical calculations. Thus our data invalidates the previously published model of magnetic structure for Fe$_4$O$_5$~[\onlinecite{ovsyannikov2016charge}], where the simultaneous magnetic ordering of all iron ions was assumed.

Among the oxides presented in Fig.~\ref{Figure_2} the~HP-Fe$_3$O$_4$ is the only non-recoverable phase and so its spectrum was acquired at 20~GPa, after in~situ synthesis in diamond anvil cell. Although HP-Fe$_3$O$_4$, as an $N=2$ member of the $N$-family, could, in principle, possess iron ions in an integer oxidation state, our data unexpectedly revealed the presence of mixed-valent states in both crystallographic positions, i.e. in the chains and slabs.

% Thus at $N\geq3$ the octahedral iron ions inevitably have a non-integer average oxidation state. We note here that convincing signs of the mixed-valent state in the chains were also observed in the charge-ordered Fe$_4$O$_5$-IV phase at pressures above~10~GPa~[\onlinecite{ovsyannikov2018pressure}]. Whether the violation of Fe$^{2+}$ preference for the trigonal prisms is a common effect of pressure in these structures or only relevant to the  HP-Fe$_3$O$_4$ is an interesting question requiring a separate study.

%\paragraph*{\textbf{Explanation of magnetism}}

\begin{figure}
\includegraphics[width=7cm, keepaspectratio=true]{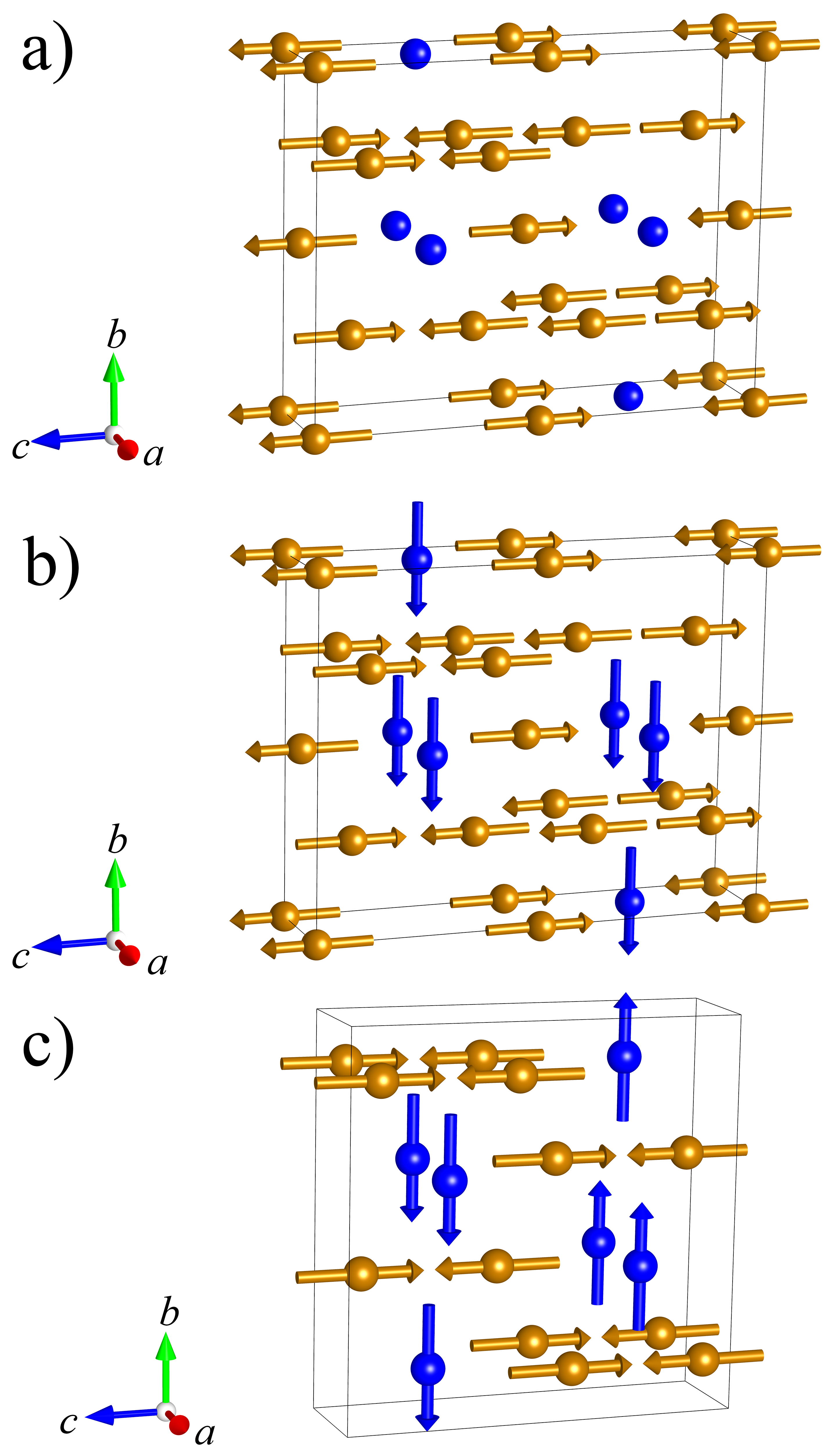}
\caption{\label{Figure_3}  The expected magnetic configurations for the charge-averaged state in the $N$-family. The transition from paramagnetic to completely magnetically ordered state proceeds via the intermediate partially ordered state~(a), with the magnetically ordered slabs but the disordered 1D chains. The complete magnetic order differs in odd and even $N$ compounds due to the ferromagnetic and antiferromagnetic \emph{interchain} ordering, respectively. We illustrate this here with (b)~$N=3$, i.e.~Fe$_4$O$_5$, and (c)~$N=2$, i.e.~HP-Fe$_3$O$_4$, structures.}
\end{figure}

To understand the origin of the independent magnetic behavior of spins in slabs and chains it is instructive to establish a generic scheme of exchange interactions in these structures. Let us consider them for Fe$_4$O$_5$, the $N=3$~member of the $N$-family, since in this case there are neutron diffraction data~\cite{hong2018CaFe3O5,hong2018MnFe3O5,hong2018CoFe3O5} for several isostructural compounds and solid solutions differing only by cations in the chains (namely Ca,~Mn or~Co), which revealed an identical collinear antiferromagnetic order of the slabs at room temperature with a (0~0~0) propagation vector presented in Fig.~\ref{Figure_3}a.

In this magnetic structure, the magnetic moments in the slabs align along the $c$-axis, with ferromagnetic exchange along the $a$-axis and antiferromagnetic exchange along the $b$- and $c$-axis~(Fig.~\ref{Figure_3}a). This pattern of exchange interactions in the octahedral slab should be relevant not only in Fe$_4$O$_5$ but can be generalized to all members of the $N$-family. This conjecture is supported by our DMFT calculations for HP-Fe$_3$O$_4$ and the data from CaFe$_5$O$_7$, which is equivalent to the $N=5$ case, where the same magnetic configuration is adopted in the ground state\cite{delacotte2017morin}. As for the chains, application of the well-known Goodenough-Kanamori-Anderson (GKA) rules suggests that they should have relatively weak \emph{intrachain} ferromagnetic coupling in case of Fe$^{2+}$ ions (see Suppl.~Materials).

\begin{figure*}
\includegraphics[width=\linewidth, keepaspectratio=true]{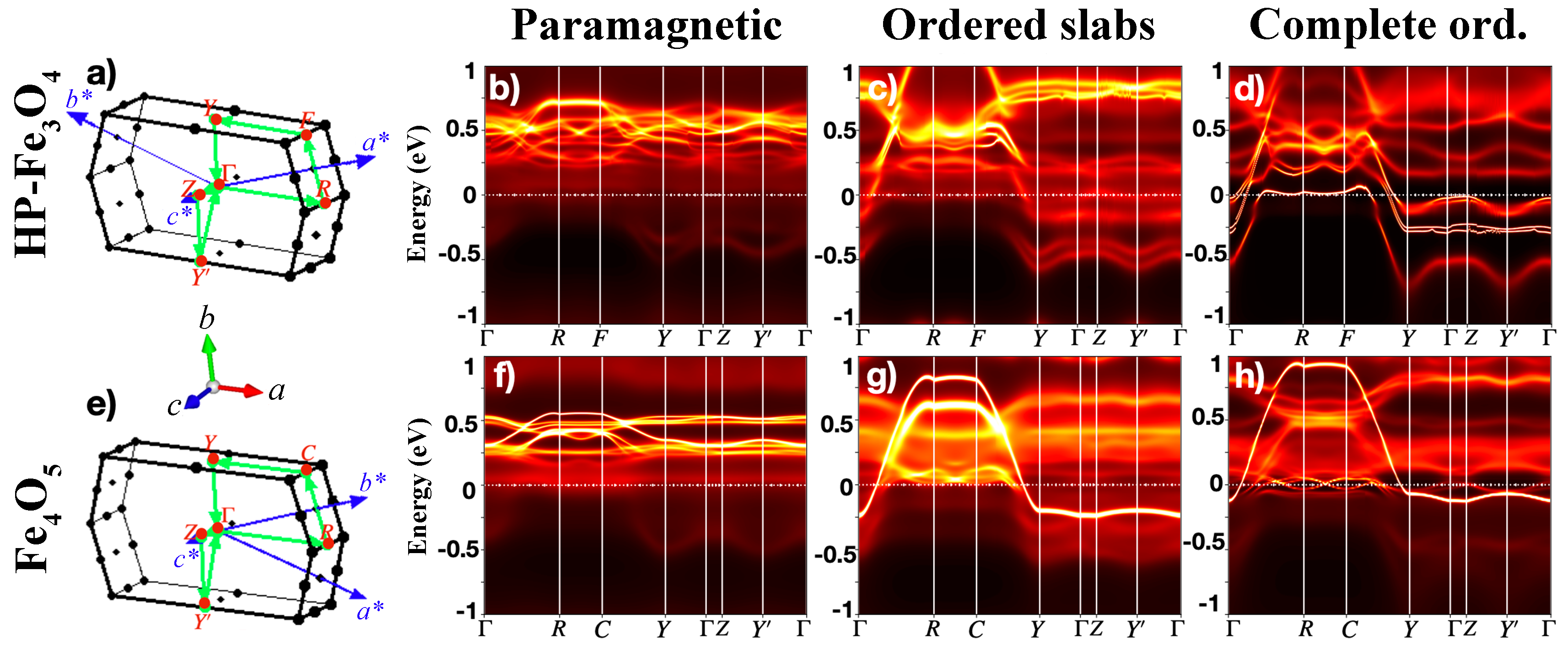}
\caption{\label{Figure_4} Theoretical eDMFT spectral functions of HP-Fe$_3$O$_4$ (top row) and Fe$_4$O$_5$ (bottom row) in three different states: paramagnetic (b and f), partially ordered state with ordered magnetic slabs and disordered chains as in Fig.~\ref{Figure_3}a (c and g) and complete long range ordering established on all Fe atoms as in Fig.~\ref{Figure_3}c and~\ref{Figure_3}b (d and h) in magnetic space groups \#63.465 and \#63.464, respectively. In a~and~e the first primitive Brillouin zones are shown with respective momentum pathway used in the plots, and between them the corresponding crystallographic axes are displayed to show the relative orientation of reciprocal space to the real space, namely $\Gamma-R$ is parallel to the $a$~axis, $\Gamma-Y$ is parallel to the $b$~axis and $\Gamma-Z$ is parallel to the $c$~axis.}
\end{figure*}

The fact that ordering in the 1D chains is independent of the ordering in the slabs is due to symmetry protection provided by the preserved twinning plane in the $N$-family. For the case of the antiferromagnetic structure in Fig.~\ref{Figure_3}a, the induced Weiss fields from magnetically ordered adjacent octahedral slabs exactly cancel out each other for iron atoms located at the mirror plane. Therefore magnetic interactions in the 1D chains are \emph{symmetry-protected} against perturbations from the octahedral slabs and also from the interchain interactions along the $c$-axis. Consequently, the chains have lower ordering temperatures, as determined by the intrachain interactions.

At intermediate temperatures this leads to the peculiar magnetic structure with coexistence of ordered slabs and disordered chains (Fig.~\ref{Figure_3}a) that is observed in our M\"{o}ssbauer spectra of Fe$_4$O$_5$ and Fe$_5$O$_6$ at room temperature in Fig.~\ref{Figure_2}. From these observations it immediately follows that such independent magnetic ordering cannot be observed in the monoclinic $N,N$+1-family. Indeed, in our DMFT simulations for Fe$_7$O$_9$ we observe non-zero Weiss field in the chains for all types of magnetic order in the slabs, contrary to Fe$_4$O$_5$ and HP-Fe$_3$O$_4$ where coexisting magnetic state (Fig.~\ref{Figure_3}a) is found to be stable, with no induced Weiss field in the chains.

Therefore upon complete magnetic ordering for the $N$-family we expect to see ferromagnetic chains embedded in the antiferromagnetic background of the slabs (Fig.~\ref{Figure_3}b and~\ref{Figure_3}c). The subtle crystallographic difference between odd and even $N$ compounds mentioned above results in ferromagnetic \emph{interchain} ordering in the odd $N$ case (Fig.~\ref{Figure_3}b) and antiferromagnetic \emph{interchain} ordering in the even $N$ case (Fig.~\ref{Figure_3}c). Thus the odd and even $N$ compounds should have a ferromagnetic and an antiferromagnetic ground state, respectively (see Suppl. Materials for more details).

%\paragraph*{\textbf{Band structure and transport properties}}

Ab-initio simulations have been performed by the combination of Density Functional Theory and embedded Dynamical Mean Field Theory (DFT+eDMFT) for two members of the $N$-family, HP-Fe$_3$O$_4$ and Fe$_4$O$_5$, i.e. even and odd $N$ case, respectively, in three different states at 120~K: (\emph{i}) the paramagnetic state with all local moments disordered, (\emph{ii}) the intermediate partially ordered state with magnetically ordered slabs but disordered 1D chains (Fig.~\ref{Figure_3}a), and (\emph{iii}) the state with complete magnetic order on all Fe atoms (Fig.~\ref{Figure_3}b and~\ref{Figure_3}c).

In accordance with the expected mixed-valent state, both compounds are metallic in all three magnetic configurations. Overall our results are consistent with previous paramagnetic (DFT+eDMFT) calculations of Fe$_4$O$_5$ and Fe$_5$O$_6$~[\onlinecite{yang2021metallic,qin2021site}] and the computed oxidation states of Fe~atoms are in good agreement with our M\"{o}ssbauer data (see occupation number $n_d$ in Extended Data Table III).

The obtained single-particle spectral functions are displayed in Fig.~\ref{Figure_4}. Fe~atoms in the slabs are coordinated by the oxygen octahedra, hence the $t_{2g}$ orbitals are lower in energy and are partially occupied with finite density of states at the Fermi level, whereas the $e_g$ orbitals are almost exactly half-filled, and are fully gapped with the large Mott gap exceeding 1~eV, see Extended Data Fig.~5~and~6. In the 1D chains, the Fe atoms are sitting in trigonal prisms (see Suppl. Materials for the orbital splitting in the trigonal prism), and here the $z^2$ and $x^2-y^2$ orbitals appear metallic, while the other three orbitals are gapped.

The divergent self-energy, seen as sharp poles in Extended Data Fig.~5~and~6, attests to the Mott insulating nature of the gapped orbitals, and shows enormous scattering rate  values for them. This implies that the Fe atoms in the paramagnetic state appear in a very unusual configuration, namely the orbitally selective Mott state~\cite{anisimov2002orbital,koga2004orbital,koga2005spin,biermann2005non,liebsch2005novel,werner2007high,de2009orbital,vojta2010orbital,huang2012pressure,herbrych2018spin,herbrych2020block,kim2022orbital}, which is well known for its bad metallic behaviour. The Mott-insulating localized electrons provide a strong source of spin scattering for the electrons in the metallic orbitals, and so the latter cannot properly conduct electrons~\cite{biermann2005non}. Perhaps the most famous example of such selective Mott states are the colossal magnetoresistive manganites~\cite{tokura1999colossal,pascut2020role}, but several other systems were recently studied as potential candidates to harbor such unusal electronic configurations~\cite{anisimov2002orbital,de2009orbital, vojta2010orbital, kim2022orbital}. Due to this large spin scattering, the spectral functions of these paramagnetic states (Fig.~\ref{Figure_4}b and~~\ref{Figure_4}f) are very diffuse and show no evidence for developed conducting bands around the Fermi level. The predicted electrical conductivity at 100~K is of the order of $10^5$~S/m and is rather isotropic (see the Extended Data Table~III).

Upon magnetic ordering of the slabs, the spin-scattering on octahedrally coordinated Fe~atoms is arrested, and so the electrons on these atoms experience very small scattering rates. The metallic orbitals now start to form bands which cross the Fermi level (see Fig.~\ref{Figure_4}c and~\ref{Figure_4}g), but these bands are still relatively fuzzy, because the spin scattering in the chains remains active. The plot of the density of states and the scattering rate in Extended data Figs.~5~and~6 confirm that not much has changed for the Fe~atoms in the chains, even though the slabs are now magnetically ordered. Consequently, the conductivity has increased by approximately a factor of two (see Extended Data Table~III), but it is still quite isotropic and small for a metallic system.

Finally the magnetic moments in the chains also order, as depicted in Fig.~\ref{Figure_3}b and~\ref{Figure_3}c, which removes all spin scattering at the Fermi level, making the system a normal magnetic metal with sharp bands near the Fermi level, as shown in Fig.~\ref{Figure_4}d and~\ref{Figure_4}h. This state shows the peculiar tendency to low-dimensional behavior, namely the strong dispersion of the bands along the $\Gamma-R$ and $F(C)-Y$ direction of the first Brillouin zone, as shown in Fig.~\ref{Figure_4}a and~\ref{Figure_4}e, while along the orthogonal directions the bands are mostly flat. The first direction is along the 1D~chains of trigonal prisms, i.e. the real space~$a$~axis, therefore we expect enhanced conductivity along the 1D~chains.

Indeed, the calculation results in Extended Data Table~III show increased anisotropic conductivity with largest values along the chains in this ordered state. In HP-Fe$_3$O$_4$ conductivity increases by an order of magnitude and the anisotropy ($2\sigma^{a}/[\sigma^b+\sigma^c]$) is about two, although we find that it is very sensitive to doping and can become as large as five upon a small shift of the Fermi level (see Extended Data Table~IV).

It is interesting to note that in the completely ordered magnetic state of HP-Fe$_3$O$_4$ the flat parts of the bands at the Fermi level are mostly formed by the $z^2$ and $x^2-y^2$~orbitals of the Fe atoms in the 1D chains, as evident by the narrow peak in the corresponding densities of states at the Fermi level (see Extended Data Fig.~5). Therefore, the chains show a clear tendency to act as an 1D conductors.

%\paragraph*{\textbf{Implications}}
Up to now we have treated these compounds only in the charge-averaged state. However, at low temperatures the inherent mixed-valence state of the slabs can act as a natural switch between metallic and insulating behavior by means of the charge-ordering transition~\cite{ovsyannikov2016charge,ovsyannikov2020room}. This means that the anisotropy could be enhanced even further in the true charge-ordered ground state, leaving the arrays of chains as the only conductive channel at low temperature.

The $N$-family is therefore a 3D system with an emergent dimensionality reduction.  This is brought about because the tropochemical cell-twinning mechanism magnetically isolates the iron ions in the trigonal prisms from the surrounding slabs, allowing the 1D nature of the chains to manifest in physical properties like the conductivity. This conclusion is rather general and this symmetry protection of magnetic interactions in the chains also applies under ionic substitution\cite{hong2018MnFe3O5,hong2018CoFe3O5} and in other examples even with the (1/2~0~0) magnetic propagation vector, as in stoichiometric CaFe$_3$O$_5$~[\onlinecite{cassidy2019single}].

Crucially, these materials are transferable to the ambient pressure realm, as distinct to the superconducting high-pressure hydrides\cite{pickard2020superconducting}. Half of the oxides listed in Extended Data Table~III are recoverable and, in addition, these structures show great compositional flexibility that considerably affects their stability field\cite{woodland2013fe,boffa2015complete,hong2019thesis}. For instance, compounds with $N\geq3$ can be synthesized \emph{even at ambient pressure} if the Fe atoms in the chains are substituted by Ca~[\onlinecite{evrard1980mise,malaman1981preparation}] while another interesting alternative is to substitute oxygen with other chalcogens, as exemplified by the lillianite case\cite{takeuchi1997tropochemical}.

Furthermore, the crystallographic derivation mechanism described here suggests that it may be possible to stabilize these compounds in the thin-film form. The inherent connection with the rock salt structure via tropochemical cell-twinning~(Fig.~\ref{Figure_1}) means that (011)-oriented substrates with rock salt structure could provide a coherent interface for the (100)-oriented films of the $N$-family members.

To conclude, the distinct magnetic and electronic properties of the structures in this series make them candidates for the realization of a large variety of interesting physical systems, particularly considering magnetoresistance. One possible implementation should be highlighted, namely the coupled quantum wire models\cite{vishwanath2001two,kane2002fractional,teo2014luttinger,iadecola2016wire,meng2015coupled}, because the arrays of conductive 1D chains described here would be an excellent testbed for them. However the range of possible phenomena is much wider and the recent observation of  the 3D quantum Hall effect in ZrTe$_5$~[\onlinecite{tang2019three}] confirm this, as it is equivalent to the $N=3$ member of $N$-family with empty cationic sites in the slabs.

\subparagraph{Acknowledgements} We are grateful to Daniel~I. Khomskii and Pavel~A. Volkov for inspiring discussions. D.V. and E.B. thank the Swedish Research Council for support under grant number 2018-04704. G.K. and K.H. acknowledge the support of National Science Foundation by the grant number DMR-1709229. We acknowledge the European Synchrotron Radiation Facility for provision of synchrotron radiation resources at the beamline ID18.
\subparagraph{Author contributions} S.O., L.D. and D.V. prepared the samples. D.V., I.K., G.A., S.C., V.C., V.P., L.D. and A.C. performed the M\"{o}ssbauer experiments. G.K. and K.H. performed the DFT+eDMFT calculations. D.V. developed the crystallographic approach. D.V., K.H. and E.B. interpreted the results and wrote the manuscript with input from all co-authors.
%\subparagraph{Competing interests} The authors declare no competing interests.

%The applied epitaxial strain can stabilize thin films of high-pressure phases and the conventional macroscopical twinning of thin film is one of the common ways to release it\cite{freund2004thin}. Would it work with the twinning on the unit cell scale remains an open question to be explored.

\section*{Methods}
\paragraph*{\textbf{Sample preparation}}
The single crystals of Fe$_4$O$_5$ and Fe$_5$O$_6$ were synthesized in a 1200-tonne multi-anvil press at the Bavarian Research Institute of Experimental Geochemistry and Geophysics (BGI) at 14-16~GPa and 1200-1400 $^{\circ}$C from the stoichiometric mixtures of  of Fe$_3$O$_4$ (Aldrich, 99.99\% purity) and Fe (99.99\%). The details of the synthesis were described in previous works\cite{ovsyannikov2018pressure,ovsyannikov2020room}. For the M\"{o}ssbauer experiments we selected single crystals using a three-circle Bruker diffractometer equipped with a SMART APEX CCD detector and a high-brilliance Rigaku rotating anode (Rotor Flex FR-D, Mo-K$_{\alpha}$ radiation) with Osmic focusing X-ray optics.

To obtain well-crystallized pure HP-Fe$_3$O$_4$ phase we performed an \emph{in situ} synthesis in a BX90 diamond anvil cell\cite{kantor2012bx90} (DAC) using laser heating at~20~GPa from a magnetite single crystal enriched with $^{57}$Fe. We used a double-sided laser-heating system with infrared lasers\cite{kupenko2012portable,aprilis2017portable}. The phase transition from spinel structure to the high-pressure phase was clearly seen from M\"{o}ssbauer spectra as at this pressure the spinel phase has two well-developed magnetic sextets while the spectrum of HP-Fe$_3$O$_4$ is a superposition of two paramagnetic doublets.

%For pressure generation we used a BX90 DAC equipped with diamonds of 250 $\mu$m culet size\cite{kantor2012bx90}. The gasket was prepared by  pre-indentation of 200-$\mu$m-thick rhenium foil to the thickness of about 30~$\mu$m and cutting of a 120-$\mu$m-diameter hole by infrared laser in the center of indentation. After placement of the sample and small ruby spheres the pressure chamber was loaded by neon or argon using gas-loading system\cite{kurnosov2008novel}. The pressure inside DAC was controlled using ruby fluorescence\cite{dewaele2008compression}.

%For measurements of Fe$_4$O$_5$ with variable temperate we equipped DAC with a miniature external platinum resistive heater\cite{dubrovinskaia2005internal,kantor2012bx90}. The sample temperature was determined using the type S thermocouple and analysis of M\"{o}ssbauer spectra, namely using temperature-dependent second-order Doppler shift of the components, see our previous work for more details about this method\cite{kupenko2019magnetism}. Since the ruby fluorescence is temperature-dependent we used Y fluorescence band of Sm-doped yttrium aluminum garnet (Sm:YAG) for the pressure control during this experiment because of its insensitivity to the temperature changes\cite{trots2013sm}.

\paragraph*{\textbf{M\"{o}ssbauer spectroscopy}}
The M\"{o}ssbauer spectra were collected at the Nuclear Resonance beamline~\cite{ruffer1996nuclear} (ID18) at the European Synchrotron Radiation Facility (ESRF) using the synchrotron M\"{o}ssbauer source (SMS)~\cite{potapkin201257fe}. The SMS is based on a high-quality single crystal of $^{57}$FeBO$_3$ which acts as a nuclear monochromator. Energy modulation is achieved by means of a standard M\"{o}ssbauer velocity transducer on which the borate crystal is mounted. The advantages of SMS are the small beam spot size, about 10--15 $\mu$m, and the absence of non-resonant radiation that greatly improves the data quality and decreases the acquisition time for experiments with DACs.

The folded M\"{o}ssbauer spectra contain 512 channels and were fitted using the MossA software~\cite{prescher2012mossa}, version 1.01a. We used a transmission integral fit assuming a Lorentzian-squared line shape of the SMS and a Lorentzian or pseudo-Voigt line shape for the absorber. The line width and the center shift of the SMS was controlled before and after collection of each spectrum using the reference single line absorber (K$_2$Mg$^{57}$Fe(CN)$_6$). All center shifts are reported relative to $\alpha$-Fe at ambient conditions.

The assignment of the oxidation state of iron ions was done based on the comparison of the determined isomer shift values with literature data of high-spin iron ions in oxygen environment\cite{menil1985systematic}. As a reference for the 6-fold oxygen coordinated iron at ambient pressure we took values of octahedral Fe$^{2.5+}$ in magnetite (0.67~mm/s) and Fe$^{3+}$ in hematite (0.36~mm/s), to minimize the influence of the inductive effect\cite{menil1985systematic}. The isomer shift of the iron ion in the +2.5 mixed-valent state is just an average of the isomer shifts of pure divalent and trivalent iron. Therefore from the magnetite value it follows that pure Fe$^{2+}$O$_6$ would have 0.98~mm/s isomer shift and this allows us to calculate the isomer shift for iron ions in any intermediate oxidation state between +2~and~+3. As can been seen from Extended Data Table~II the iron ions in the trigonal prisms of Fe$_4$O$_5$, Fe$_5$O$_6$ and Fe$_7$O$_9$ oxides have isomer shifts very close to our estimate, confirming occupation of the trigonal prismatic sites by pure Fe$^{2+}$. Under this assumption we can calculate the average oxidation state of iron ions in the octahedral slabs and in the same manner predict isomer shift values for them.  In Extended data Table~II one can see a remarkable agreement between experimental and predicted values.

The spectrum of HP-Fe$_3$O$_4$ was collected after synthesis at 20~GPa so we need to make a correction for the pressure dependence of isomer shift which we thoroughly studied in our previous work\cite{Vasiukov_spin_transition}. Although from the stoichiometry of HP-Fe$_3$O$_4$ one would expect only pure Fe$^{2+}$ and Fe$^{3+}$ in the chains and slabs, respectively, the determined isomer shift values are quite far away from the expected numbers (Extended Data Table~II). Note that the deviation from the calculated number is opposite for the chains and slabs, which corresponds to an admixture of trivalent iron to the chains and divalent iron to the slabs. Therefore the mixed-valent state is adapted by both crystallographic positions in this phase. The results of our DMFT calculations in Extended Data Table~III also show a similar trend.

\paragraph*{\textbf{DFT+eDMFT calculations}}

The theoretical calculations were carried out by the combination of DFT and eDMFT~\cite{kotliar2006electronic}, as implemented in \texttt{WIEN2k} and the Rutgers eDMFT code~\cite{blaha2020wien2k,haule2010dynamical,haule2018structural}. These calculations are charge self-consistent and use exact double-counting between DFT and DMFT~\cite{haule2015exact}. For the self-consistent calculation we set $RK_{max}$ (which determines the size of basis) to be 7.0 and used 1000 $k$-points for the $k$-point sampling in the irreducible Brillouin zone. To compute the resistivity, we used up to 10000 $k$-points. We adopted the local-density approximation (LDA)~\cite{perdew1992accurate}, which gives the best results for lattice properties when combined with eDMFT~\cite{haule2015free}. The self-energy on the real axis was obtained using the analytical continuation by maximum entropy method. The experimental crystal structures used in calculations were taken from works~[\onlinecite{lavina2011discovery,bykova2016structural}], namely Fe$_4$O$_5$ at 10~GPa and HP-Fe$_3$O$_4$ at 44~GPa. We used $U$=10.0 eV and $J_H=1.0$ eV for the Coulomb repulsion and Hund's coupling respectively, as obtained by the constrained DMFT calculation. The auxiliary impurity problem was solved using a continuous-time quantum Monte Carlo impurity solver~\cite{haule2007quantum}.

The magnetic structures were first guessed by finding the Weiss field on each Fe atom in the presence of the exchange field of the neighboring Fe atoms, which can provide several candidate orderings which can be further simulated for their stability.
Among several trial configurations, we found that the theoretical ordering of Fe magnetic moments, depicted in Fig.~\ref{Figure_3}, is indeed energetically favourable to other configurations.

Note that the spin configurations used in calculations (Extended Data Figs.~7~and~8) are non-collinear, and canted along the high-symmetry axis of the octahedra, which minimizes the sign problem in quantum Monte Carlo. Namely, in the DMFT calculation the local coordinate axis is chosen such that it coincides with the high symmetry axis of the polyhedron, which minimizes the non-diagonal hybridization, and hence reduces the sign problem. The spin-orientation along the $z$-axis then does not introduce additional off-diagonal hybridization, which makes the sign problem in magnetic calculation approximately equal to the paramagnetic case.

Extended data Table~III shows the Fe occupation numbers~$n_d$. In this work, the DMFT calculations were carried out in the complete Linear Augmented Plane Wave basis, and the occupation numbers were computed by integrating the electronic charge within the touching Muffin-tin spheres around iron atoms and projecting them to the relevant $d$~orbitals.

The theoretical conductivity in the same table was calculated as the zero frequency limit of the optical conductivity, which is computed as the convolution of the single-particle Green's functions with the proper velocity matrix elements. Notice that the vertex corrections to conductivity vanish within the single site eDMFT framework because considered here structure have the inversion symmetry.  All visualizations of crystal and magnetic structures in this work were prepared using the VESTA software\cite{momma2011vesta}.

\bibliography{Bibliography/My_bib}

\clearpage
\renewcommand{\figurename}{Extended data Fig.}
\renewcommand{\tablename}{Extended data Table}

\begin{sidewaysfigure*}
\includegraphics[width=\linewidth, keepaspectratio=true]{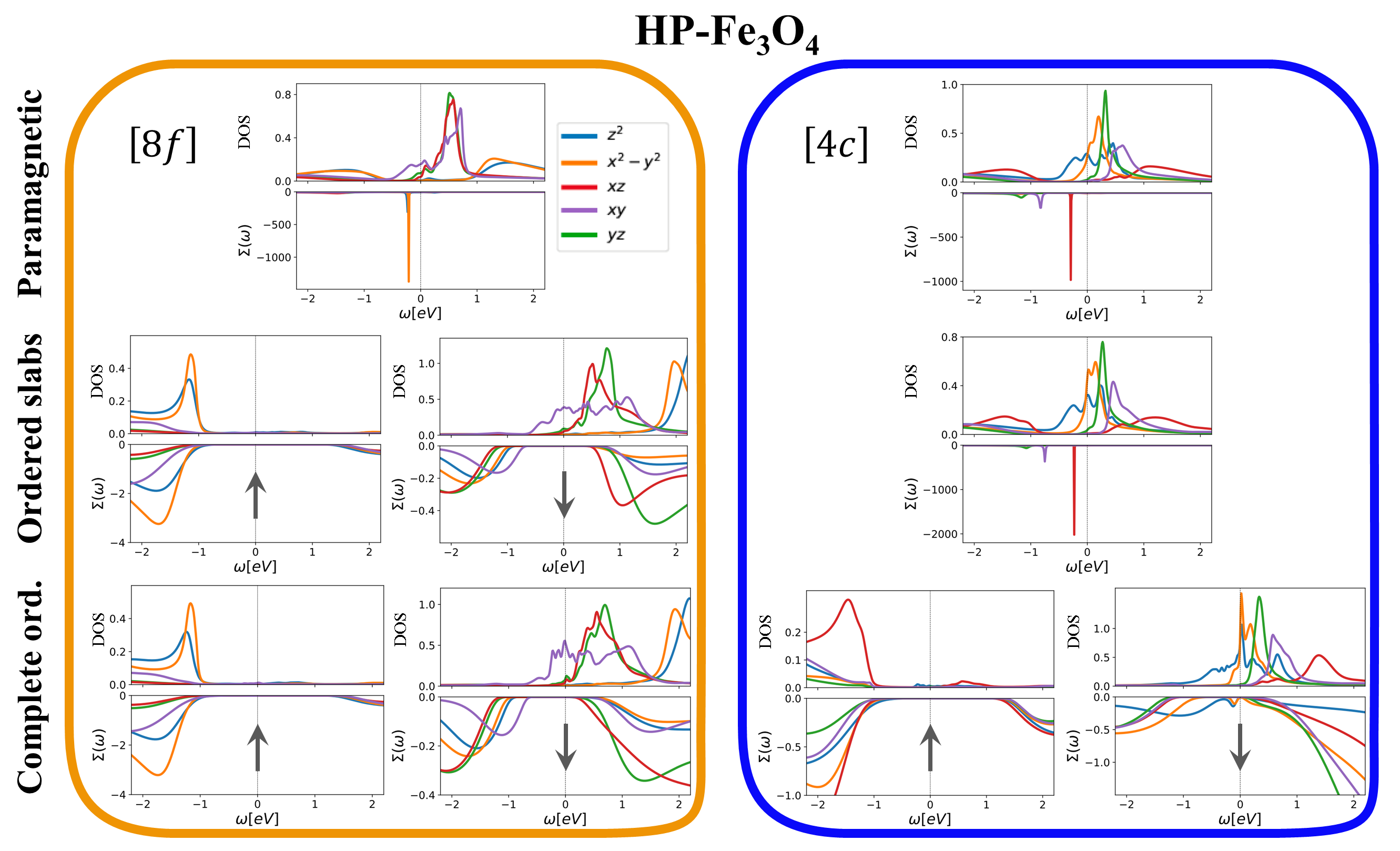}
\caption{\label{HP-Fe3O4_DOS} Theoretical eDMFT electronic density of states (DOS) and self-energy in HP-Fe$_3$O$_4$ for three different states: the paramagnetic state (top row),  the intermediate state with ordered slabs and disordered chains (middle row) and the completely ordered state (bottom row). For each state we show the projected electronic DOS on the Fe~$3d$~orbitals in the upper panel, while the lower panel displays the imaginary part of the self-energy, which gives the electronic scattering rate. The plots for the slabs ($8f$~site) and the chains ($4c$~site) are contoured with orange and blue frames, respectively. The up and down arrows mark the spin-majority and spin-minority states, respectively.}
\end{sidewaysfigure*}

\begin{sidewaysfigure*}
\includegraphics[width=\linewidth, keepaspectratio=true]{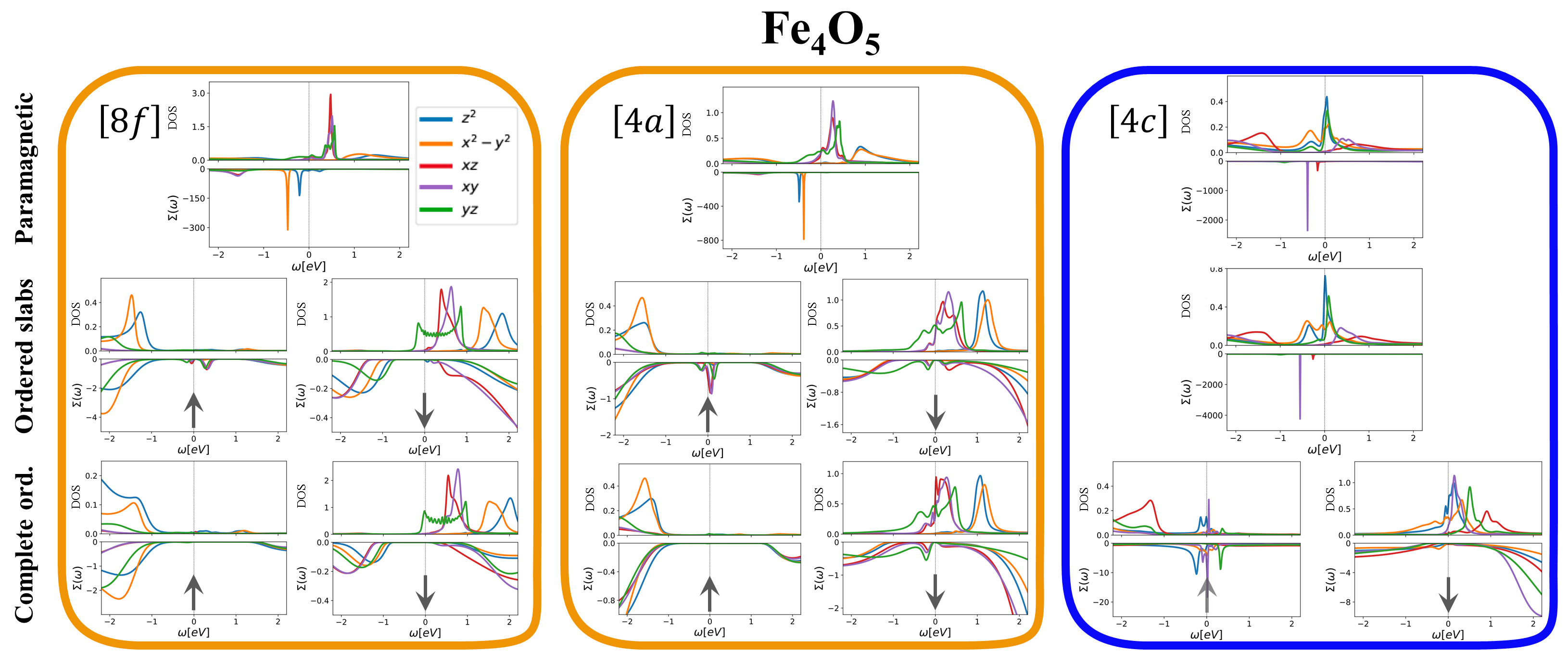}
\caption{\label{Fe4O5_DOS} Theoretical eDMFT electronic density of states (DOS) and self-energy in Fe$_4$O$_5$ for three different states: the paramagnetic state (top row),  the intermediate state with ordered slabs and disordered chains (middle row) and the completely ordered state (bottom row). For each state we show the projected electronic DOS on the Fe~$3d$~orbitals in the upper panel, while the lower panel displays the imaginary part of the self-energy, which gives the electronic scattering rate. The plots for the slabs ($8f$~and $4a$~sites) and the chains ($4c$~site) are contoured with orange and blue frames, respectively. The up and down arrows mark the spin-majority and spin-minority states, respectively.}
\end{sidewaysfigure*}

\begin{figure*}
\includegraphics[width=\linewidth, keepaspectratio=true]{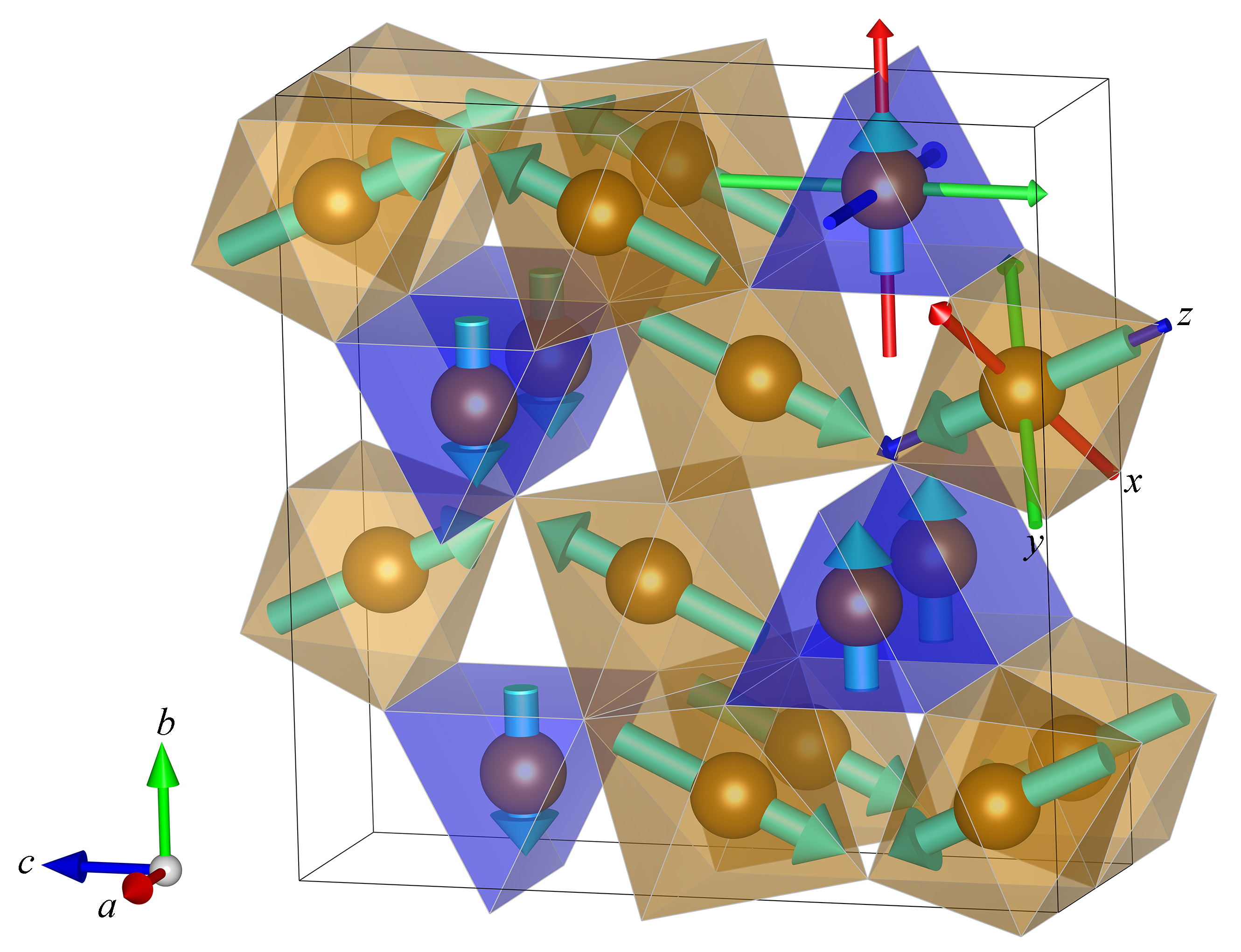}
\caption{\label{Fe3O4_DMFT_CS} The canted magnetic structure and local coordinate systems used for the eDMFT calculations of HP-Fe$_3$O$_4$. The red, green and dark blue axes correspond to the local $x$, $y$~and $z$~axes, respectively.}
\end{figure*}

\begin{figure*}
\includegraphics[width=\linewidth, keepaspectratio=true]{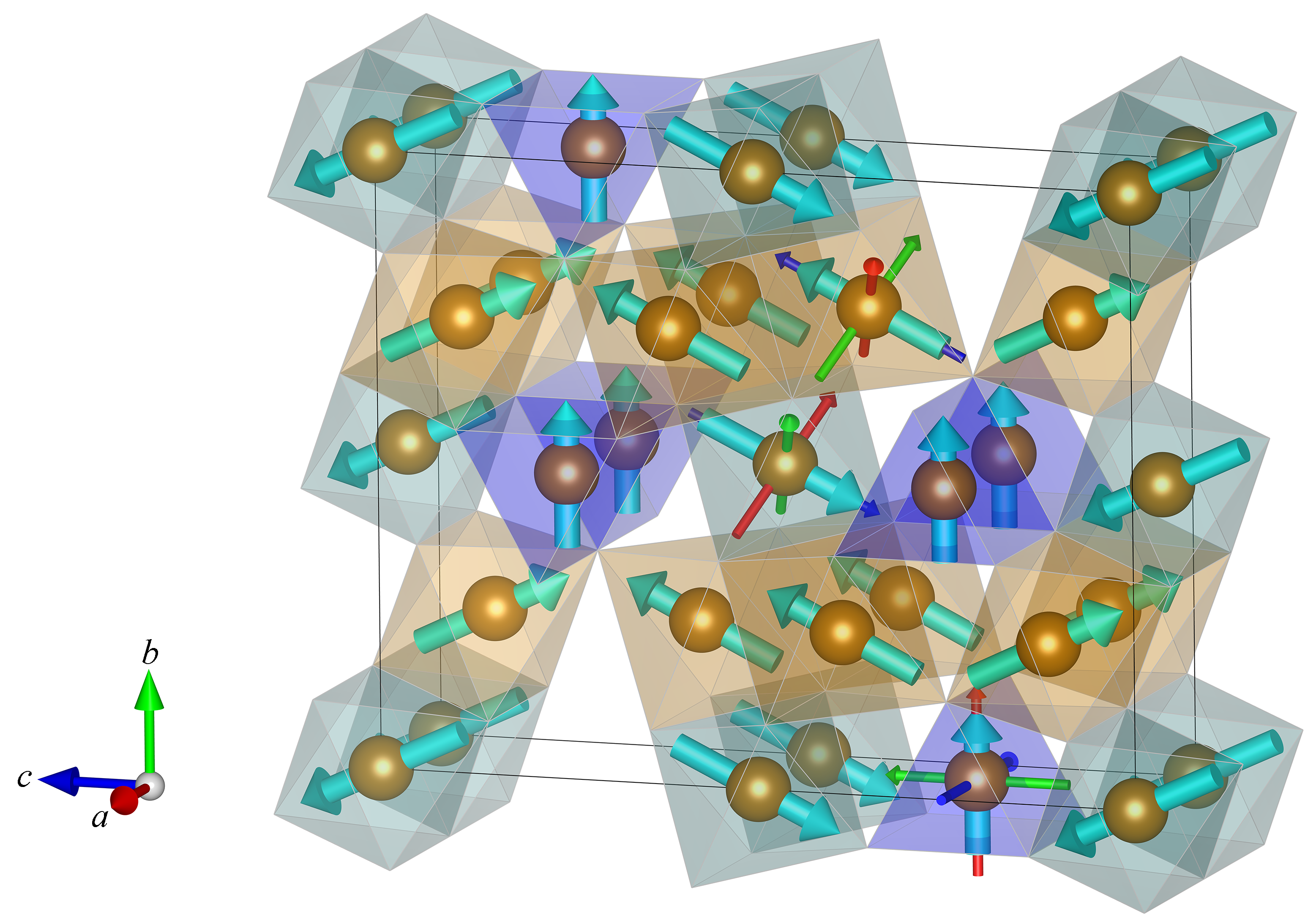}
\caption{\label{Fe4O5_DMFT_CS} The canted magnetic structure and local coordinate systems used for the eDMFT calculations of Fe$_4$O$_5$. The red, green and dark blue axes correspond to to the local $x$, $y$~and $z$~axes, respectively, in accord with the previous figure. The light green octahedra correspond to Fe atoms at the $4a$~Wyckoff site.}
\end{figure*}

\begin{turnpage}
\begin{table}[h]
\caption{\label{Cryst_data} Crystallographic data of the $n$FeO$\cdot m$Fe$_2$O$_3$ homologous series. Among the currently known oxides of this series, two subseries can be distinguished with different type of the slab cycle: an orthorhombic $N$-family with space group $Cmcm$ and a monoclinic $N$,$N$+1-family with space group $C2/m$. The recently discovered Fe$_7$O$_{10}$ has a more complicated slab cycle. The relevant data of the parent w\"{u}stite structure (FeO) are also provided for comparison. The $a$ and $b$ lattice constants of the $N$-family are equivalent to the $b$ and $a$ lattice constants of the $N$,$N$+1-family, respectively.}
\begin{ruledtabular}
\begin{tabular}{cccccccccccc}
Compound & SG & Slab cycle & \multicolumn{2}{c}{Iron sites} & P, GPa & \emph{a}, \AA & \emph{b}, \AA & \multicolumn{2}{c}{\emph{c}, \AA} & $\beta$ & Ref.\\
\hline
$N$-family & & & \textcolor{blue}{Tr. prism} & \textcolor{orange}{Octahedron} & & & & exp. & calc. & &\\
\hline
$\eta$-Fe$_2$O$_3$ & $Cmcm$ & 1 & \textcolor{blue}{4c(m2m)} & \textcolor{orange}{4a(2/m..)} & 64 & 2.640 & 8.639 & 6.414 & & & [\onlinecite{bykova2016structural}]\\
HP-Fe$_3$O$_4$ & $Cmcm$ & 2 & \textcolor{blue}{4c(m2m)} & \textcolor{orange}{8f(m..)} & 44 & 2.694 & 9.282 & 9.309 & & & [\onlinecite{bykova2016structural}]\\
Fe$_4$O$_5$ & $Cmcm$ & 3 & \textcolor{blue}{4c(m2m)} & \textcolor{orange}{4a(2/m..) $|$ 8f(m..)} & ambient & 2.891 & 9.802 & 12.580 & 12.965 & & [\onlinecite{ovsyannikov2016charge}]\\
Fe$_5$O$_6$ & $Cmcm$ & 4 & \textcolor{blue}{4c(m2m)} & \textcolor{orange}{8f(m..) $|$ 8f(m..)} & ambient & 2.877 & 9.917 & 15.340 & 15.558 & & [\onlinecite{ovsyannikov2020room}]\\
\hline
$N$,$N$+1-family& & & & & & \emph{b}, \AA & \emph{a}, \AA & & & \\
\hline
Fe$_5$O$_7$ & $C2/m$ & 1,2 & \textcolor{blue}{4i(m)} & \textcolor{orange}{2a(2/m) $|$ 4i(m)} & 41 & 2.733 & 9.208 & 8.270 & & $105.50^{\circ}$ & [\onlinecite{bykova2016structural}]\\
Fe$_7$O$_9$ & $C2/m$ & 2,3 & \textcolor{blue}{4i(m)} & \textcolor{orange}{2d(2/m) $|$ 4i(m) $|$ 4i(m)} & ambient & 2.895 & 9.696 & 11.428 & 11.916 & $101.69^{\circ}$ & [\onlinecite{sinmyo2016discovery}]\\
\footnote{Fe$_9$O$_{11}$ was stabilized as a solid solution with Mg and its actual composition is Mg$_{0.9}$Fe$_{8.1}$O$_{11}$~[\onlinecite{ishii2018synthesis}].}Fe$_9$O$_{11}$ & $C2/m$ & 3,4 & \textcolor{blue}{4i(m)} & \textcolor{orange}{2a(2/m) $|$ 4i(m) $|$ 4i(m) $|$ 4i(m)} & ambient & 2.892 & 9.844 & 14.176 & 14.480 & $99.956^{\circ}$ & [\onlinecite{ishii2018synthesis}]\\
\hline
& & & & & & \emph{a}, \AA & \emph{b}, \AA & & & \\
\hline
Fe$_7$O$_{10}$ & $Cmcm$ & 1,1,2 & \textcolor{blue}{4c(m2m) $|$ 8f(m..)} & \textcolor{orange}{8f(m..) $|$ 8f(m..)} & 64 & 2.652 & 8.767 & 21.96 & & & [\onlinecite{koemets2021chemical}]\\
Fe$_{0.94}$O & $Fm\bar{3}m$ & $\infty$ & & & ambient & \footnote{The periodicity along \textless110\textgreater.}3.041 & \footnote{The periodicity along \textless332\textgreater.}10.084 & \footnote{The actual cubic lattice constant.}4.3 & & & [\onlinecite{foster1956metal}]\\
\end{tabular}
\end{ruledtabular}
\end{table}
\end{turnpage}

\begin{table*}[h]
\caption{\label{Hyperfine_param}Parameters of M\"{o}ssbauer spectra of the studied high-pressure iron oxides at room temperature. $\delta_{CS}$ is the center shift, $\Delta E_Q$ is the quadrupole splitting, $\varepsilon$ is the quadrupole shift and $B_{hf}$ is the hyperfine magnetic field. The expected octahedral Fe$^{2+}$:Fe$^{3+}$ ratio and calculated center shift values (see Methods section for details) are evaluated assuming that the trigonal prismatic site is occupied only by Fe$^{2+}$.}
\begin{ruledtabular}
\begin{tabular}{ccccccccc}
Compound & \multicolumn{2}{c}{\textbf{HP-Fe$_3$O$_4$}} & \multicolumn{2}{c}{\textbf{Fe$_4$O$_5$}} & \multicolumn{2}{c}{\textbf{Fe$_5$O$_6$}} & \multicolumn{2}{c}{\textbf{Fe$_7$O$_9$}}\\
Tr. prism/oct. & \multicolumn{2}{c}{\textcolor{blue}{1}:\textcolor{orange}{2}} & \multicolumn{2}{c}{\textcolor{blue}{1}:\textcolor{orange}{3}} & \multicolumn{2}{c}{\textcolor{blue}{1}:\textcolor{orange}{4}} & \multicolumn{2}{c}{\textcolor{blue}{2}:\textcolor{orange}{5}}\\
\parbox[c][1cm]{2.7cm}{\linespread{1}\selectfont{Expected oct. Fe$^{2+}$/Fe$^{3+}$ ratio}} & \multicolumn{2}{c}{\textcolor{orange}{0:1}} & \multicolumn{2}{c}{\textcolor{orange}{1:2}} & \multicolumn{2}{c}{\textcolor{orange}{1:1}} & \multicolumn{2}{c}{\textcolor{orange}{1:4}}\\
Pressure & \multicolumn{2}{c}{20(1) GPa} & \multicolumn{2}{c}{ambient} & \multicolumn{2}{c}{ambient} & \multicolumn{2}{c}{ambient}\\
& \textcolor{blue}{Tr. prism} & \textcolor{orange}{Oct.} & \textcolor{blue}{Tr. prism} & \textcolor{orange}{Oct.} & \textcolor{blue}{Tr. prism} & \textcolor{orange}{Oct.} & \textcolor{blue}{Tr. prism} & \textcolor{orange}{Oct.}\\
\hline
expt. $\delta_{CS}$, mm/s & \textcolor{blue}{0.73(2)} & \textcolor{orange}{0.45(2)} & \textcolor{blue}{0.98(2)} & \textcolor{orange}{0.59(1)} & \textcolor{blue}{0.97(3)} & \textcolor{orange}{0.70(1)} & \textcolor{blue}{0.954(3)} & \textcolor{orange}{0.561(3)}\\
calc. $\delta_{CS}$, mm/s & \textcolor{blue}{$0.90\pm0.04$} & \textcolor{orange}{$0.30\pm0.02$} & \textcolor{blue}{0.98} & \textcolor{orange}{0.57} & \textcolor{blue}{0.98} & \textcolor{orange}{0.67} & \textcolor{blue}{0.98} & \textcolor{orange}{0.48}\\
$\Delta E_Q$/$2\varepsilon$, mm/s & \textcolor{blue}{2.32(4)} & \textcolor{orange}{$\sim0$} & \textcolor{blue}{2.34(4)} & \textcolor{orange}{0.44(1)} & \textcolor{blue}{1.08(4)} & \textcolor{orange}{0.67(1)} & \textcolor{blue}{2.736(6)} & \textcolor{orange}{0.348(4)}\\
$B_{hf}$, T & \textcolor{blue}{0} & \textcolor{orange}{0} & \textcolor{blue}{0} & \textcolor{orange}{24.78(5)} & \textcolor{blue}{0} & \textcolor{orange}{32.87(4)} & \textcolor{blue}{0} & \textcolor{orange}{0}\\
Area, \% & \textcolor{blue}{36(5)} & \textcolor{orange}{64(5)} & \textcolor{blue}{21(4)} & \textcolor{orange}{79(4)} & \textcolor{blue}{19(2)} & \textcolor{orange}{81(2)} & \textcolor{blue}{26.9(6)} & \textcolor{orange}{73.1(6)}\\
%$\Gamma$, mm/s & \textcolor{blue}{0.2(1)} & \textcolor{orange}{0.34(7)} & \textcolor{blue}{0.44(10)} & \textcolor{orange}{0.50(4)} & \textcolor{blue}{0.49(12)} & \textcolor{orange}{0.45(4)} & \textcolor{blue}{0.10(1)} & \textcolor{orange}{0.22(2)}\\
\end{tabular}
\end{ruledtabular}
\end{table*}
\clearpage

\begin{table*}[t]
\caption{\label{table_Mag} Theoretical eDMFT electric conductivity $\sigma$ at $100\,$K, occupation $n_d$ and magnetic moment $M$ for each Fe-site in HP-Fe$_3$O$_4$ and Fe$_4$O$_5$. Here $\sigma^b$ and $\sigma^c$ are the conductivities along the $b$ and $c$~axes, respectively, while $\sigma^a$ corresponds to the $a$~axis, i.e.~along the 1D chains. The slabs of Fe$_4$O$_5$ have another Fe-site at the $4a$~Wyckoff position, in addition to the $8f$~position.}
        \begin{ruledtabular}
        \begin{tabular}{c|c|ccc|c|c}
         Compound & Magnetic order & \multicolumn{3}{c}{$\sigma$, $10^5\cdot$S/m} & Fe occupation $n_d$& Magn. moment, $\mu_B$\\
         & & $\sigma^a$ & $\sigma^b$ & $\sigma^c$ & (\textcolor{blue}{4c}, \textcolor{orange}{8f}, \textcolor{orange}{[4a]}) & (\textcolor{blue}{4c}, \textcolor{orange}{8f}, \textcolor{orange}{[4a]})\\
          \hline
                         & \emph{Complete ordering} & 55 &  17  & 25 & (\textcolor{blue}{5.80}, \textcolor{orange}{5.58}) & (\textcolor{blue}{4.12}, \textcolor{orange}{4.28})\\
          HP-Fe$_3$O$_4$ & \emph{Ordered slabs}  & 2.4 & 2.0 & 2.3 & (\textcolor{blue}{5.80}, \textcolor{orange}{5.57}) & (\textcolor{blue}{0}, \textcolor{orange}{4.28}) \\
                         & \emph{Paramagnetic}   & 1.2 & 1.1 & 0.9 & (\textcolor{blue}{5.80}, \textcolor{orange}{5.57}) & (\textcolor{blue}{0}, \textcolor{orange}{0})\\
          \hline
                      & \emph{Complete ordering} & 5.0 & 3.0 & 3.0 & (\textcolor{blue}{5.83}, \textcolor{orange}{5.80}, \textcolor{orange}{5.51}) & (\textcolor{blue}{3.44}, \textcolor{orange}{4.08}, \textcolor{orange}{4.4})\\
          Fe$_4$O$_5$ & \emph{Ordered slabs}     & 2.2 & 2.3 & 2.4 & (\textcolor{blue}{5.90}, \textcolor{orange}{5.76}, \textcolor{orange}{5.58}) & (\textcolor{blue}{0}, \textcolor{orange}{4.06}, \textcolor{orange}{4.34})\\
                      & \emph{Paramagnetic}      & 1.9 & 1.5 & 1.6 & (\textcolor{blue}{5.92}, \textcolor{orange}{5.72}, \textcolor{orange}{5.61}) & (\textcolor{blue}{0}, \textcolor{orange}{0}, \textcolor{orange}{0})\\
        \end{tabular}
        \end{ruledtabular}
\end{table*}
\clearpage

\begin{table*}[h]
\caption{\label{FE_shift} Variations of the conductivity and its anisotropy in HP-Fe$_3$O$_4$ on electronic doping in the completely ordered magnetic state. The discrepancy in conductivity values with Extended Data Table~III is due to the use of a less dense mesh in these calculations.}
\begin{ruledtabular}
\begin{tabular}{cccc}
Doping, electron per formula unit & \multicolumn{3}{c}{Conductivity, S/m}\\
 & $\sigma^a$ & $\sigma^b$ & $\sigma^c$\\
\hline
$+0.09$ & 30.6 & 16.1 & 16.4\\
$+0.06$ & 115.0 & 37.1 & 34.8\\
0 & 43.5 & 15.0 & 28.5\\
$-0.05$ & 15.0 & 5.2 & 6.3\\
$-0.07$ & 25.6 & 5.9 & 5.4\\
\end{tabular}
\end{ruledtabular}
\end{table*}

\end{document}